\documentclass[useAMS,usenatbib,referee]{mn2e}
\usepackage{epsfig}
\usepackage{graphicx}
\usepackage{amssymb}

\begin{document} 

\title[Binaries and NS velocities]
{Pulsar spin-velocity alignment from single and binary neutron star progenitors}

\author[A.G. Kuranov, S.B. Popov, K.A. Postnov]
{A.G. Kuranov $^{1}$, S.B. Popov $^{1}$, K.A. Postnov $^{1,2}$
\thanks{E-mail: alex@xray.sai.msu.ru(AGK); polar@sai.msu.ru (SBP)}\\
$^1${\sl Sternberg Astronomical Institute, Universitetski pr. 13,   
Moscow, 119991, Russia} \\
$^2${\sl Department of Physics, Moscow State University, 119992, Russia}\\}
\date{Accepted ......  Received ......; in original form ......
      %(MNRAS, 2005, in press)
      }

\maketitle

\begin{abstract}
The role
of binary progenitors of neutron stars in the apparent
distribution of space velocities and spin-velocity alignment observed 
in young pulsars is studied. A Monte-Carlo synthesis of
pulsar population from single and binary stars with different assumptions 
about the NS natal kick model (direction distribution, amplitude, and
kick reduction 
in binary progenitors which experienced mass exchange due to Roche
lobe overflow with initial masses on the main sequence from the range 
8-11 $M_\odot$) is performed. The calculated spin-velocity alignment 
distributions are compared with observational data obtained from radio polarization measurements. 
The observed space velocity of pulsars is found to be mostly shaped
by the natal kick velocity form and its amplitude; 
 the fraction of binaries
is not important here for reasonably large kicks. 
The distribution of kick direction relative to the spin axis 
during the formation of a NS 
is found to  affect strongly the spin-velocity correlation of pulsars.
Comparison with observed pulsar spin-velocity angles 
favours a sizeable fraction of binary progenitors and the kick-spin
angle $\sim 5-20^\circ$.
The form of the initial binary mass ratio
distribution does not affect our results.
\end{abstract}

\begin{keywords}
stars: neutron --- pulsars: general --- X-rays: binaries
\end{keywords}

\section{Introduction}

Neutron stars (NS) resulting from core collapses of massive stars 
are observed in very different astrophysical sources (radio pulsars,
compact central objects in supernova remnants, soft gamma-ray repeaters,
anomalous X-ray pulsars, cooling radio-quiet NSs, X-ray binaries, etc.), 
and new observational appearences of NSs 
are discovered nearly every few years 
(rotating radio transients are one of the latest examples;
for a comprehensive review on recent studies in this field
see, e.g., the conference proceedings \citealt{40yrs}).
Determination of the birth properties of NSs is crucial 
to understand the physics of supernovae and to model the evolution of NSs.
For a given equation of state of the NS matter, main physical parameters of a newborn NS which determine its evolution and
observational appearences  include its 
mass, spin period, surface magnetic field, and the initial space velocity. 
The situation is more complicated when a NS is formed in a binary 
system. Then the subsequent evolution of the NS will be 
also determined by 
the binary orbital parameters and the evolutionary state
of the secondary component prior to the supernova explosion, 
in addition to initial conditions of the 
collapse. 

In single radio pulsars, NS space velocities (more precisely, 
their tangent component) can be measured directly. 
Using complicated models, one can reconstruct the distribution of 
the NS birth velocities 
(see, for example, \cite{fk2006} and references therein). 
 The shape of the resulting distribution is still ambiguous. 
Some authors \citep{acc2002} favour a bimodal Maxwellian distribution, 
some \citep{hobbs2005} argue that the data can be explained with a single 
Maxwellian curve, 
while others \citep{fk2006} prefer single-mode non-Maxwellian distributions. 
Moreover, the space velocity (both in magnitude and direction) 
can be correlated with other parameters. 
In pulsars, one of the best known correlations  
is found between the direction of the space velocity vector 
and the spin axis of the NS.
It is well established for the Crab and Vela pulsars 
and is found in several other objects with well studied pulsar wind nebulae 
\citep{nr2004,nr2007}. Polarimetric measurements of radio pulses 
also  allow to estimate 
 the angle $\delta$ between the spin axis and the NS space velocity direction  
\citep{j2005,r2007}. 
The spin-velocity correlation is firmly established only in young pulsars, 
since for older ones the motion in 
the galactic gravitational potential smears out any 
initial relation between the NS spin and velocity vectors.

Recently, \cite{j2007} reported new measurements of the angle  $\delta$
between the spin axis and space velocity direction in 14 radio pulsars. 
Half of them  demonstrate strong alignment, 
while others do not. 
New pulsar measurements and observations of pulsar wind nebulae are expected to 
increase the list of objects with known $\delta$ (e.g. \citealt{r2007}).

The origin of the spin-velocity alignment is usually related to the kick mechanism 
operating during or immediately after the stellar core collapse. 
The reason why NSs obtain so large 
(up to $>$~1000~km~s$^{-1}$) space  velocities remains unknown. 
Several absolutely different physical mechanisms are discussed 
(see, for example, the table in \citealt{bp2004}, 
and reviews in \citealt{lcc2001} and \citealt{pyu07}).

Historically, the first proposed kick mechanism was related to an  
asymmetric mass ejection during the supernova explosion 
\citep{s1970}. Recently, a modification of
this mechanism was studied in numerical
simulations by \cite{setal2004,setal2006}.
In these simulations large kicks can be recovered, but the direction of the 
velocity is found 
to be poorly correlated with the NS spin axis. 
Nevertheless, in the framework of this mechanism
the correlation is possible for  small spin periods of newborn NSs 
($P<1$~msec), as rapid rotation results in averaging of momentum along all
directions except the spin vector \citep{lcc2001}, and anisotropies appear
not randomly but preferably along the rotation axis \citep{setal2006}. 

In some models, the NS kick is related to an 
asymmetric neutrino emission in a strong magnetic field. The first
scenario of this type was proposed by  \cite{c1984}
(see also the paper by \citealt{BK1993}). 
Several variants of this mechanism were later discussed by \cite{lcc2001}.
In the neutrino scenarios, the spin-velocity alignment is possible if 
the neutrino emission time is longer than the spin period or if 
the magnetic dipole is initially aligned with the spin axis.

A tight spin-velocity alignment naturally emerges
in two other mechanisms: 
when the kick is due to asymmetry between two oppositely directed matter jets
emanating from a newborn NS \citep{k1999,a2003}, or when a 
rotating NS with asymmetric (e.g. off-centered)  magnetic dipole 
is gradually accelerated by electromagnetic radiation 
(the electromagnetic rocket mechanism suggested by \citealt{ht1975}).
The spin-velocity alignment may be a consequence of the magneto-rotational supernova explosion \citep{abkm05}.

A perfect orthogonal spin-velocity misalignment is expected in the case
of the collapsing core fragmentation due to rapid rotation
 \citep{b1988, i1992, in1992, cw2002}.
However, it is unclear whether this mechanism can be 
generic for the majority of NSs, 
since numerical models of stellar evolution suggest  
rather slow rotation of the stellar core prior to the collapse
\citep{heger&2005}.  

If all NSs were originated from isolated progenitors, 
deviations from the spin-velocity alignment 
could be related to differences in the kick mechanism 
or in the NS behavior immediately after the formation
(the typical space velocity of massive stars -- NS-progenitors, $\sim 10-30$~km s$^{-1}$,  
is much smaller than the NS kick velocities
and so can not  significantly affect the 
relative orientation of the spin and
velocity axis, see the results below). 
This posibility was studied, for example, by \cite{nr2007}.
However, at least a half of isolated NSs are expected to be born 
in binaries so the net space velocity acquired by a NS formed in a binary system should be 
the sum of the NS natal kick and 
the orbital velocity of the NS progenitor prior to the binary disruption
(plus the velocity of the binary system barycenter). 
Even if the kick mechanism generates a single-mode velocity distribution, 
the additional contribution from progenitor's orbital motion
in the binary system  
can change its shape. While the effects of the binary progenitor
on the observed NS space velocity distribution can be noticeable only  
for small kicks (of order 
or smaller than the orbital velocity of the collapsing star, typically 
$<100$~km~s$^{-1}$), 
they can appear more pronounced in the NS spin-velocity misalignment. 
In our previous paper \citep{postnov&kuranov2008} we studied the 
NS spin-kick velocity correlation effect on binary 
NS coalescence rates and spin-orbit misalignment of the components.
In the present paper we specially address the role of binaries in 
shaping the NS space velocity 
distribution and their effect on the spin-velocity alignment in 
observed radio pulsars.

\newpage

\section{Population synthesis code}

 For our population synthesis calculations we use the {\rm SCENARIO MACHINE}
code, developed at Sternberg Astronomical Institute (see 
\cite{lpp1996,lppb2007} and references therein\footnote{See the
on-line material at
http://xray.sai.msu.ru/$\sim$mystery/articles/review/.}).
The general description of the population synthesis techniques can be found
in \cite{pp2007}.

The standard assumptions about the
binary evolution are made: the Salpeter mass function for the primary mass,
$dN/dM_1\sim M_1^{-2.35}$,
the initial semi-major axis distribution 
in \"Opik's form $dN/d\log a={\rm const}$, 
$$
\label{loga}
 ~~~~~~ \max \left\{
\begin{array}{lcl}10R_\odot\\
\rm{R_L(M_1)}\\
\end{array}
\right\}
<a<10^7  R_\odot,
$$
where $R_{\mathrm L}(M_1)$ is the radius of the Roche lobe of the primary component.

Calculations are done for two assumptions about
the initial mass ratio ($q=M_2/M_1<1$) distribution: 
$dN/dq\sim{q^0}$ 
(flat distribution), and the mass ratio distribution 
$dN/dq\sim{q^2}$ which is strongly peaked  towards equal masses of
the components. The latter choice is related to
recent studies by \cite{ps2006} who found a large fraction of twins
among 21 eclipsing  
massive binaries in SMC. The real initial distribution 
of binary mass ratio in our galaxy is still unknown, but 
the choice of these two limiting cases (the flat one and the one 
strongly 
peaked towards equal masses) gives us  a flavor how sensitive are
our results to this important initial parameter.

The common envelope phase of the binary evolution is treated in
the standard way based on the energy conservation \citep{pyu07} with
the efficiency $\alpha_\mathrm{CE}=0.5$.

We assume that the kick velocity vector 
is confined within a cone 
coaxial with the progenitor's rotation axis 
and  characterized by the angle
$\theta \le \pi/2$. 
We shall consider only central kicks thus ignoring
theoretically feasible off-center kicks simultaneously affecting the NS
spin~\citep{Spruit,Postnov&Prokhorov,Wang_ea07}. 

\begin{table}
\caption{Model of the kick}
\bigskip
\begin{center}
\begin{tabular}{lccr}
\hline
Model & NS progenitors & $v_\mathrm{p}$ km~s$^{-1}$ & Initial mass \\
\hline
       &         & &\\
\bf{SA}& single  & 100 ... 500 & $> 10M_\odot$ \\
       &         & & \\
\hline
       &         & & \\
\bf{BA}& binary  & 100 ... 500 & $> 10M_\odot$\\
       &         & & \\
\hline
       &         & & \\
       &         & 30  &   $8M_\odot$--$11 M_\odot$\\
\bf{BB}& binary  & 100...500 & $> 11M_\odot$\\
       &         & & \\
\hline
       &         & & \\
\bf{BAS}       &50$\%$ binary  & 100...500&  $> 10M_\odot$\\
\bf{(BA + SA)}       &   +      & & \\
          &50$\%$ single & 100...500& $> 10M_\odot$ \\
       &         & & \\
\hline
       &         & & \\
      &      & 30 & 8$M_\odot$ --$11M_\odot$\\
\bf{BBS}       & 50$\%$ binary & 100...500  & $> 11 M_\odot$\\
\bf{(BB + SA)}        &    +     & & \\
        &50$\%$ single & 100...500 & $> 10M_\odot$ \\
       &         & & \\
\hline
\label{t_model}
\end{tabular}
\end{center}
\end{table}

We made two assumptions about the NS kicks.

(a) Kick type A. It is a single-Maxwellian (for the sake of simplicity)
distribution $f(v)\sim (v^2/v_\mathrm{p}^3) \exp(-(v/v_\mathrm{p})^2)$, 
as suggested by pulsar proper motion
measurements \citep{hobbs2005}. The parameter $v_\mathrm{p}$ is 
varied within the range 100-500 km s$^{-1}$.
For NSs from single progenitors
we always assume this type of the kick. 

(b) Kick type B. It assumes
kick type A for NSs from single progenitors and from binary 
progenitors with maximum masses during the lifetime above $11 M_\odot$.
For NSs from binary progenitors
with the maximum lifetime masses in the range 8-11$M_\odot$
that experienced the Roche
lobe overflow, 
in which  
the electron-capture core collapse can occur 
(\citealt{petal2004, vdH04, vdH07}), 
$v_\mathrm{p}$ is reduced and kept fixed at 30~km~s$^{-1}$.

According to the type of the kick assumed, we made calculations
for different model populations of NSs, as summarized in Table \ref{t_model}.
Model SA means kick type A from single progenitors, 
BA means kick type A from binary progenitors, 
BB means kick type B from binary progenitors, 
BAS means that calculations are performed for equal
number of single and binary progenitors with kick type A assumed, 
and BBS means that calculations are performed for equal
number of single progenitors with kick type A 
and binary progenitors with kick type B.

The kick type B appears to be
more realistic and most of the results below are given 
for this model of the kick.
 It is necessary to note, that some fraction of NSs originated from
isolated progenitor can be formed after e$^-$-capture SN. \cite{poletal2008}
estimate this fraction as 4\%, however it can be slightly higher, because
stellar physics prior to SN is not well understood, yet. \cite{vdH07} notes
that the fraction of normal radio pulsars originated from e$^-$-capture SNae
cannot be high, because there is no room for large fraction of low-velocity
NSs among pulsars. On the other hand, one can speculaes that such NSs do not
manifest themslves as normal pulsars. They can ``hide'' among radio silent
compact objects. For example, close-by cooling NSs (aka ``The Magnificent
Seven'') and central compact objects in supernova remnants. Product of
e$^-$-capture SNae are expected to be light. Light NSs cool slower (see a
review in \citep{yp04}). 
And those types of radio silent
NSs are know to be hot sources (see a brief review in \citealt{popov2008}).
Anyway, in this study we neglect the possibility that NS from initially
isolated progenitors can be born after e$^-$-capture SN.

For both isolated and binary stars 
we also add some center-of-mass velocity, 
typical for massive stars
in the galactic disc. For this velocity we use the 
Maxwellian distribution with
$v_\mathrm{p}=10$~km~s$^{-1}$. 
The results of our calculations 
are found to be not very sensitive to the exact value of this
parameter to within a factor of $\sim 2$. 
For a detailed study of space velocities of isolated massive
stars see \cite{l1979, s1991}.

The rotational axes of both components are assumed to
be aligned with the orbital angular momentum before the primary collapse forming the first NS in the system.
The supernova explosion is treated in a standard 
way in terms of an instantaneous
loss of mass of the exploding star.
The effect of the kick on the post-explosion 
binary orbital parameters is 
treated using the energy-momentum conservation 
in the two point-mass body problem
(see the description in e.g. \citealt{Hills83,Kalogera,Grishchuk}). 
The first supernova most likely occurs when the binary orbit is circular 
(unless the initial binary is very wide so that tidal circularization is ineffective), 
while the second explosion in the binary can
happen before the orbit has been tidally circularized.
Possible mass transfer phases before the second collapse
(such as the common envelope stage and
mass accretion onto the neutron star due to the Roche lobe overflow at 
the periastron passages) 
are assumed to effectively circularize the orbit. 
In the absence of mass transfer the tidal evolution of the orbital  eccentricity 
is treated according to \cite{Zahn_77}.
When treating the SN explosion in an eccentric
binary, the orbital position of the explosion is chosen 
randomly with a uniform distribution over the mean anomaly.

\newpage

\section{Results: the role of binaries}

In this section we discuss the effect of binary
progenitors of NSs on the pulsar space velocities and spin-velocity  
misalignment. To this purpose, we calculate populations of
isolated NSs arising from purely single and binary progenitors
(models SA, BB), and
 populations resulted from equal number 
of single and binary progenitor 
stars (models BAS, BBS).

\subsection{Space velocity of young NSs}

First we consider the NS velocity distributions 
calculated from isolated
and binary progenitors. 
An isolated NS originating in a binary system acquires both
the natal kick velocity and the orbital velocity of the exploding 
star immediately before the collapse. 
The result is sensitive to the mutual orientation of the kick and orbital velocities. 
In Fig.~1 we plot space velocities of NSs calculated for single and binary
progenitors 
in comparison with a purely Maxwellian 
velocity distribution. As expected, the influence
of the orbital velocity in binary progenitors 
is insignificant for large kick velocities
($v_\mathrm{p}\sim 300$ km~s$^{-1}$, as was adopted in this plot). 
However, 
for kick type B, where a noticeable fraction 
of NSs in binaries recieved a reduced kick, the emerging  
NS velocity distribution slightly deviates from the pure Maxwellian law
at low velocities. It appears quite difficult to check observationally such a
small effect of e$^-$-capture SN on the velocity distribution. 

\begin{figure}
\includegraphics[width=320pt,angle=0]{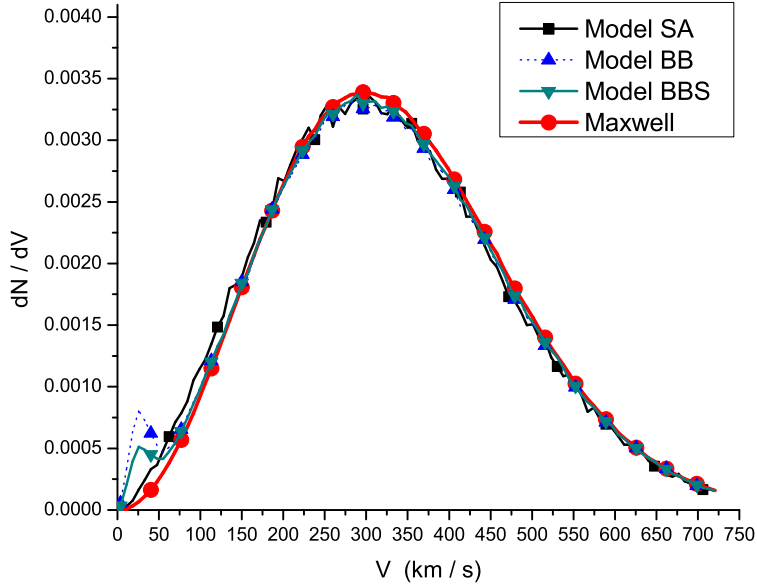}
\caption{The space velocity distribution of isolated young NSs (pulsars).
The results are shown for the model SA (black squares), 
model BB (black triangles) and model BBS (grey upside-down triangles), 
with 
$v_\mathrm{p}=300~$~km~s$^{-1}$. 
A pure Maxwellian distribution (filled circles) 
is plotted for comparison. Note a small deviation 
from the pure Maxwellian
distribution at low velocities for models BB and BBS.} 
\label{profile}
\end{figure}   

\subsection{Spin-velocity alignment in young NSs}

\begin{figure}
\includegraphics[width=320pt,angle=0]{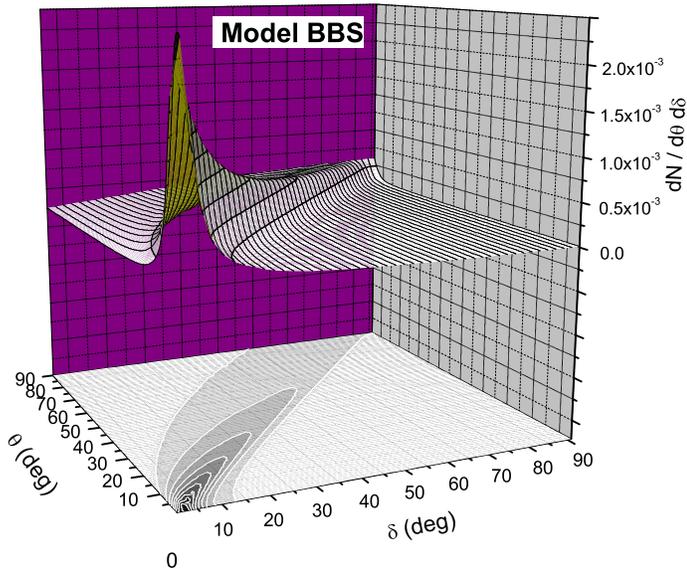}    
\caption{The spin-velocity alignment effect (the distribution of
angles $\delta$ between the spin axis and space velocity of 
a pulsar) 
for different kick-spin angles $\theta$. Calculations are made for the
model BBS with $v_\mathrm{p}=300$~km~s$^{-1}$, $dN/dq\sim{q^2}$.
The isolines on the plot stand for 10 equal intervals of the distribution 
$dN/d\theta/d\delta$ from 0 to the maximum value $2.4\cdot 10^{-3}$. }
\label{delta}
\end{figure}

The effect of binary progenitors is more interesting
for the expected NS spin-velocity alignment.
In Fig. \ref{delta} we plot the calculated spin-velocity  
alignment angle $\delta$ for different kick-spin angles $\theta$. 
The results are shown for young 
NSs in the model BBS. 
The spin and velocity vectors appear more aligned for a 
strong initial kick-spin alignment 
(i.e. small angles $\theta$). 

\begin{figure}
\includegraphics[width=340pt,angle=0]{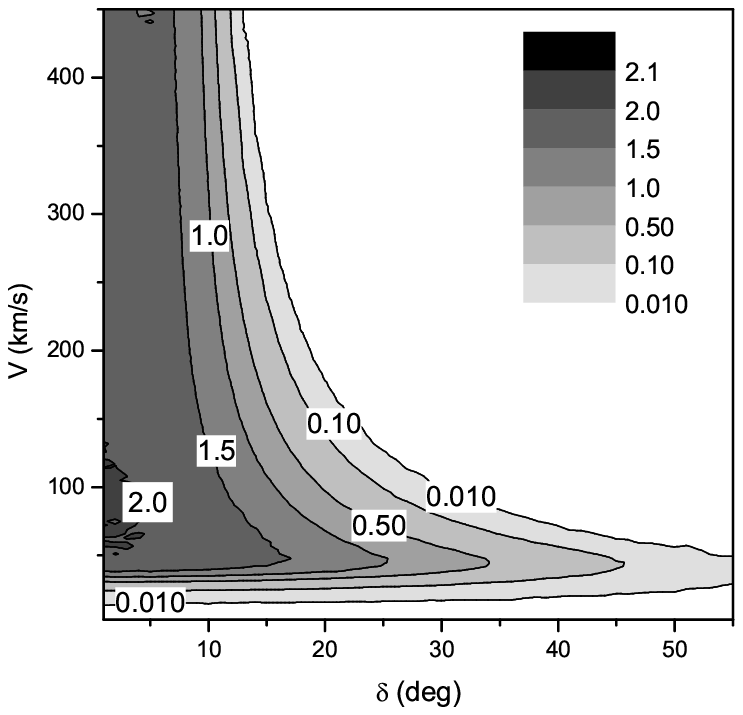}
\caption{The ratio of 
single and binary progenitors  in the
population of isolated NSs with different $\delta$ and $v$. 
The calculations are shown in terms of N(model SA)/N(model BB)
with 
$v_\mathrm{p}=300$~km~s$^{-1}$ and a 
narrow kick-spin angle $\theta=8^\circ$, $dN/dq\sim{q^2}$.
{The values $>1$ correspond to domination of single progenitors.} 
It is seen that NSs from isolated progenitors are more abundant at small
velocities $v$ (except very low values $\la 30$~km~s$^{-1}$) 
and angles $\delta$.
% (the ratio of single to binary progenitors is more than one). 
In contrast, pulsars with  $v\ga100$~km~s$^{-1}$ and
$\delta \ga 15^\circ$~--~$20^\circ$ are more likely 
to be born in a binary than
from a single progenitor (the single/binary progenitor ratio is smaller than one).
} 
\label{new_d}
\end{figure}   

The effect of binaries 
can be also illustrated (see Fig. \ref{new_d}) 
showing the relative contribution 
of single and binary progenitors to the pulsars with given $v$ and $\delta$.
Calculations are made for the models SA and BB 
with $v_\mathrm{p}=300$~km~s$^{-1}$,
 and a fiducial value of the kick-spin alignment of 
$\theta=8^\circ$ (the justification is given in Section 4). 
%  and 50/50 mixture of single/binary progenitors. 
It is seen that NSs from single progenitors (model SA)  
are more abundant (i.e. the ratio of single to binary progenitors
is more than one) at low velocities $v$ 
(except for very low values $\la 30$~km~s$^{-1}$)  
and small angles $\delta$. In contrast, pulsars with  $v\ga100$~km~s$^{-1}$ and
$\delta \ga 15^\circ$~--~$20^\circ$ are more likely 
to be born in a binary than
from a single star (the single/binary progenitor ratio is smaller than one).

\begin{figure}
\includegraphics[width=380pt,angle=0]{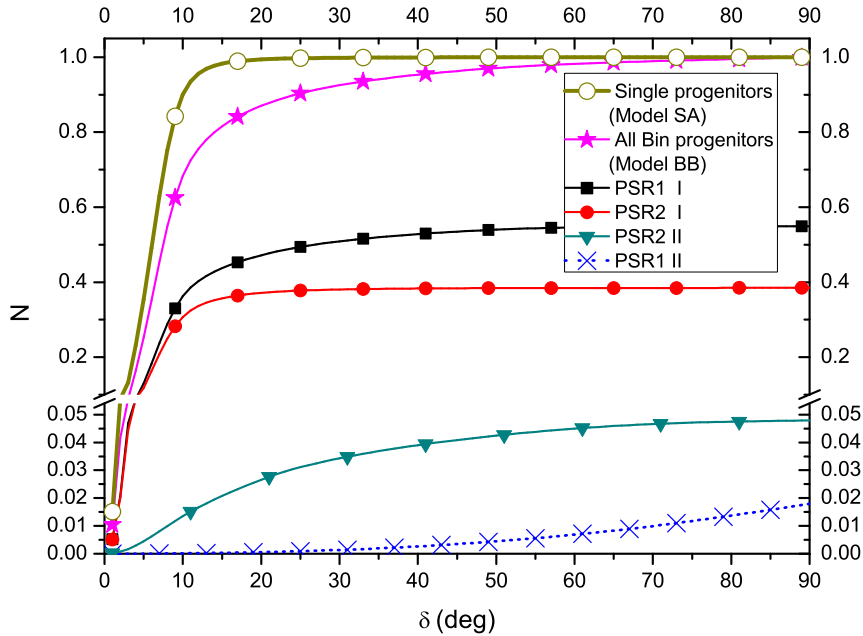} 
\caption{The cumulative distribution of the spin-velocity 
alignment angles $\delta$ for different subtypes of
pulsar progenitors. From bottom to top:
pulsars from primary components in binaries  
disrupted  after the
second explosion ($PSR$1~II, crosses);
pulsars from secondary components in binaries 
disrupted after the second explosion ($PSR$2~II, triangles);
pulsars from  secondary components in binaries  
disrupted after the 
first explosion ($PSR$2~I, filled circles); 
pulsars from  primary components in binaries 
disrupted  after the first SN explosion ($PSR$1~I, squares).
Stars show the total population of pulsars from all binary 
progenitors.
Pulsars from single progenitors are shown by open circles.
The calculations are shown for 
model SA and model BB with $v_\mathrm{p}=300$~km~s$^{-1}$ and
$\theta=8^\circ$, $dN/dq\sim{q^2}$.} 
\label{psrtypes}
\end{figure}

Let us consider more closely the spin-velocity alignment in pulsars 
originated from different progenitors (Fig. \ref{psrtypes}). 
A pulsar can result either from a single massive star or 
from the primary or secondary component 
of a massive binary system. In pulsars from single stars the spin-velocity
alignment should reflect the kick-spin correlation during the NS 
formation. In pulsars from binary progenitors the initial correlation can 
be different depending on the
 conditions and history of the NS formation. The NS formed after the collapse 
of the primary component $M_1$ (NS1) can become an isolated radio pulsar
($M_1\to NS1=PSR1$~I) immediately after the first supernova explosion if the binary 
becomes unbound, or after the second supernova explosion 
($M_1\to NS1=PSR1$~II) (filled squares and crosses
in Fig. \ref{psrtypes}, respectively).
Pulsars can also result from the collapse of the secondary 
component $M_2$ when 
the binary system has already been disrupted by  the first 
supernova explosion ($M_2\to NS2=PSR2$~I) or when the system became unbound 
after the second supernova ($M_2\to NS2=PSR2$~II).
These cases are marked in Fig. \ref{psrtypes} with filled circles and triangles, 
respectively. If the
system remained bound after both explosions, a binary NS is formed;
we discussed their coalescences in 
our previous paper \citep{postnov&kuranov2008}. 
For comparison, open circles in Fig. \ref{psrtypes} show the spin-velocity angle distribution for 
pulsars originated from single progenitors. 

Fig. \ref{psrtypes} demonstrates that 
the population of single pulsars originated from binaries 
is largely dominated by those born after 
the disruption of the parent binary  
system by the first supernova  
($PSR1$~I and $PSR2$~I), 
which constitute roughly $\sim 50\%$ and $\sim 40\%$ of 
the entire population, respectively. 

The effect of the e$^-$-capture SNs,  which is taken into acount in the
models BB and BBS, should be mentioned explicitly.
$PSR1$~I  in these models  are NSs originated from stars
with $M_1>11$~M$_\odot$, i.e. those which did not
experienced an e$^-$-capture SN.  Conditions for an e$^-$-capture SN are
such that a system always survives after the first explosion. 
$PSR2$~II are mostly NSs from systems where the first
explosion was an e$^-$-capture 
event presumably producing  small kicks so that the binary
system survives, while
the second explosion was due to the usual core
collapse event. This happens when the secondary companion significantly
increased its mass above 11 $M_\odot$ due to mass transfer 
from the primary companion.
 
Pulsars originated from binaries disrupted
after the second supernova ($PSR1$~II and  $PSR2$~II) 
provide a minor contribution but have, on
average, larger spin-velocity alignment angles $\delta$. 
Note that recycled pulsars, which can be spun-up 
by accretion in the binary after the first SN explosion, are among 
our population $PSR1$~II and constitutes less than few per cent of all
pulsars.

\newpage

\section{Discussion}

\begin{figure}
\includegraphics[width=350pt,angle=0]{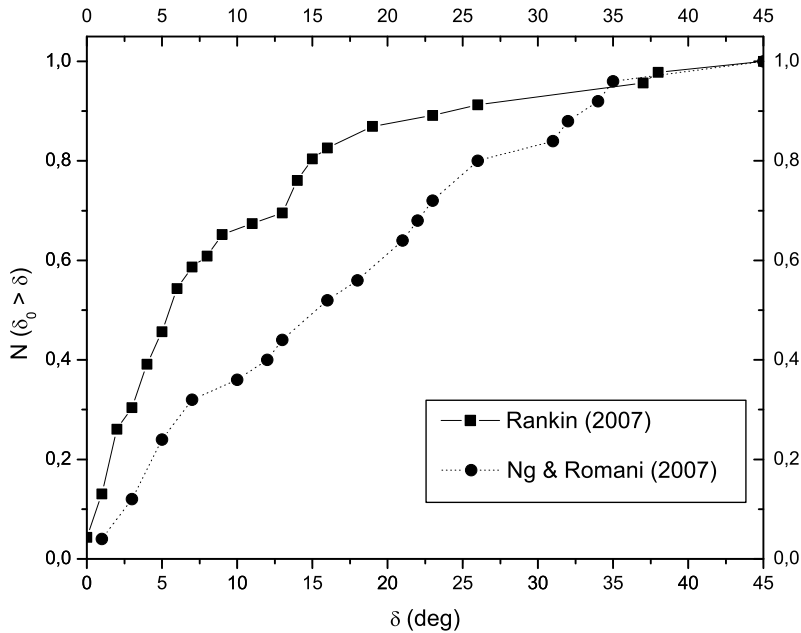}
  \caption{The cumulative distribution of the observed spin-velocity 
angles in pulsars folded around 45$^\circ$ 
due to the 90$^\circ$ ambiguity in alignment angles measured by 
radio polarization (data from Rankin 2007, filled squares, and 
Ng \& Romani 2007, filled circles).}
\label{obs_bin}
\end{figure}

It is interesting to compare the calculated spin-velocity angles
$\delta$ in single pulsars with observations assuming different model parameters.
Unfortunately,
there are only a few precise measurements. 
In \cite{nr2007} and \cite{r2007}
the authors present lists of pulsars for which the angle
$\delta$ can be inferred from observations. 
The cumulative (normalized) 
distributions of spin-velocity alignment angles (folded around 45$^\circ$ 
due to the 90$^\circ$ ambiguity in alignment angles measured by 
radio polarization) in pulsars  
from these lists are shown in  Fig. \ref{obs_bin}.
For several 
well-studied objects (see \citealt{nr2007} and references therein)
this angle can be determined from a detailed 3D-modeling 
with account of the X-ray data about the pulsar wind nebula  shape, etc. 
For most pulsars the existing estimates of the spin-velocity 
angle $\delta$ 
are based only on radio observations. In the last case
the angle between the pulsar spin axis 
and its space velocity 
is derived from measurements of the proper motions 
of the pulsar on the sky and the polarization
angle of radio emission.  
The angle determined as a difference between 
the proper motion direction
and the pulse polarization angle spans the range from $\sim
-60^\circ$ to $\sim 100^\circ$ \citep{j2007,r2007}.
However, as we are interested 
just in the angle between the velocity and rotation
axis, irrespective of the velocity direction, 
this angle should lie between 0 and 90 degrees. 
In addition, projection effects complicate 
the interpretation of data
\citep{j2005}.
On the other hand, for several well-studied objects, for example Vela \citep{nr2007}, 
it is known 
that although the difference between the proper motion direction and polarization
angle points to the spin axis perpendicular to the velocity,  
in reality $\delta$ is close to zero. This is
interpreted as being due to 
the pulsar predominantly emitting radiation
which is polarized perpendicular to the 
magnetic field lines. So it is
quite common \citep{nr2007,j2007} to fold data in such a way that
the value of the spin-velocity angle 
lies within the range  $0^\circ<\delta<45^\circ$. 

\begin{figure}
\includegraphics[width=220pt,angle=0]{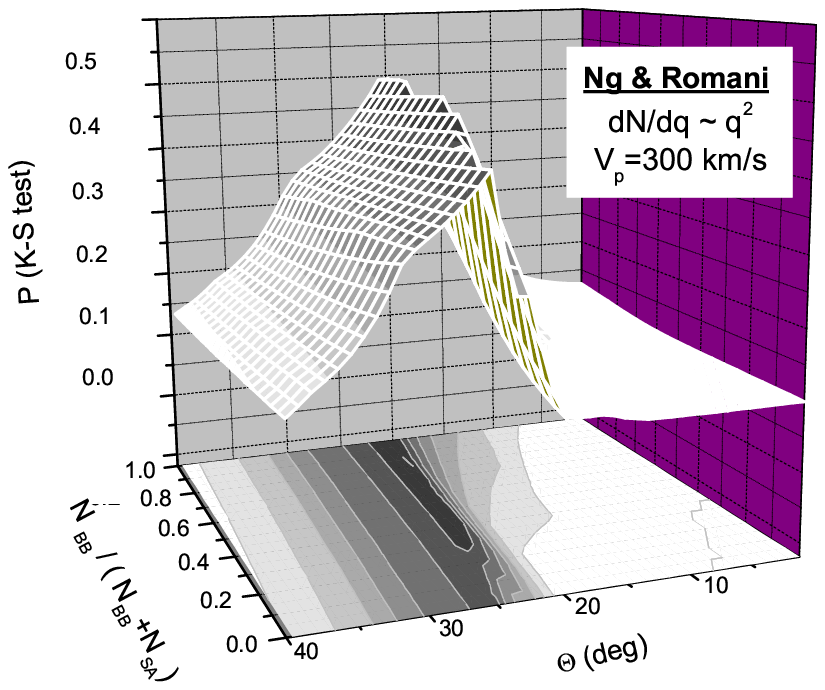}
\includegraphics[width=220pt,angle=0]{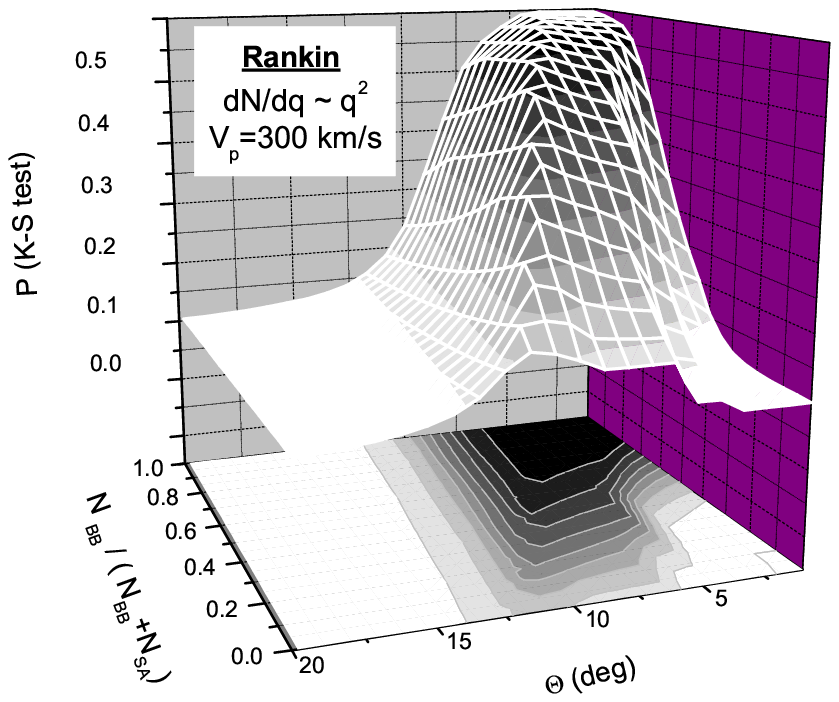}
\caption{The  Kolmogorov-Smirnov test for the observed
and calculated spin-velocity alignment angles $\delta$ in pulsars
as a function of the assumed kick-spin alignment angle $\theta$ 
and the ratio of 
binary and single progenitors of the
population of isolated NSs. The latter is given
as N(model BB)/[N(model BB)+N(model SA)].
Calculations are shown for the model BBS. The initial binary
mass ratio distribution is $f(q)\sim q^2$,
and the kick velocity amplitude is equal to 
300 km~s$^{-1}$.
Left panel: comparison with the data from Ng \& Romani (2007).
Right panel: comparison with the data from Rankin (2007). 
}
\label{ks3d_bin_s}
\end{figure}

\begin{figure}
\includegraphics[width=400pt,angle=0]{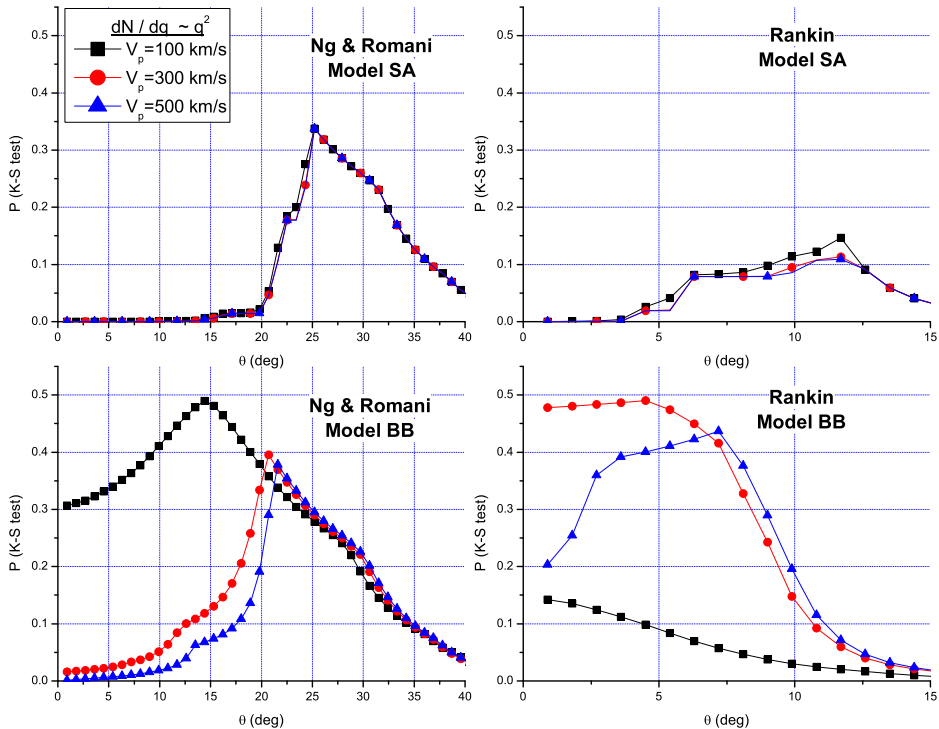}
\caption{The  Kolmogorov-Smirnov test for the observed
and calculated spin-velocity alignment angles $\delta$ in pulsars
as a function of the assumed kick-spin alignment angle $\theta$.
 Calculations are shown for the initial binary
mass ratio distribution $f(q)\sim q^2$
and models SA and BB.
Squares, circles and triangles mark the kick velocity amplitudes
100, 300 and 500 km~s$^{-1}$, respectively.
Left panels: comparison with the data from Ng \& Romani (2007).
Right panels: comparison with the
data from Rankin (2007). 
}
\label{ks_bin_s}
\end{figure}

\begin{figure}
\includegraphics[width=400pt,angle=0]{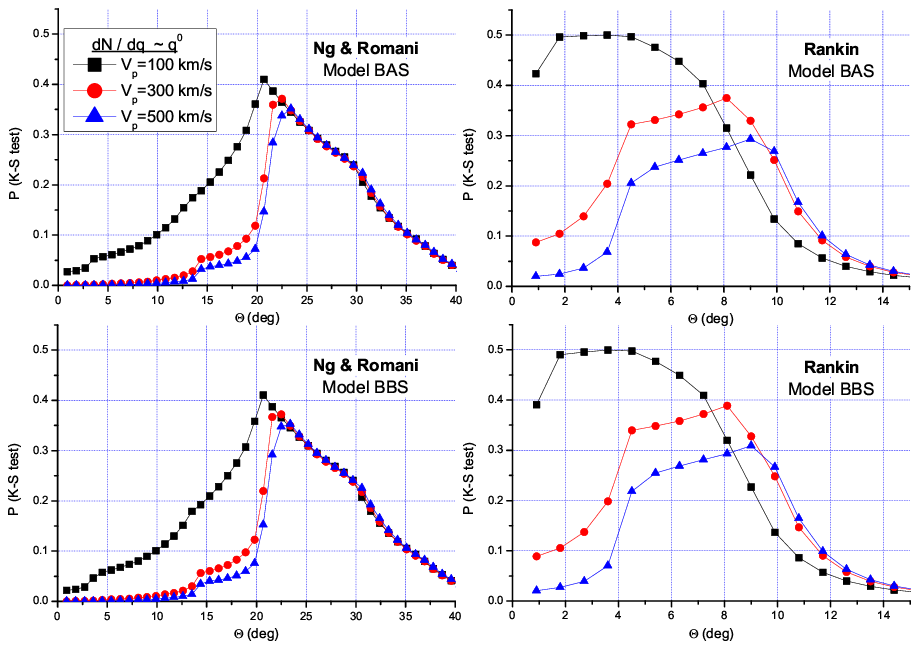}
\includegraphics[width=400pt,angle=0]{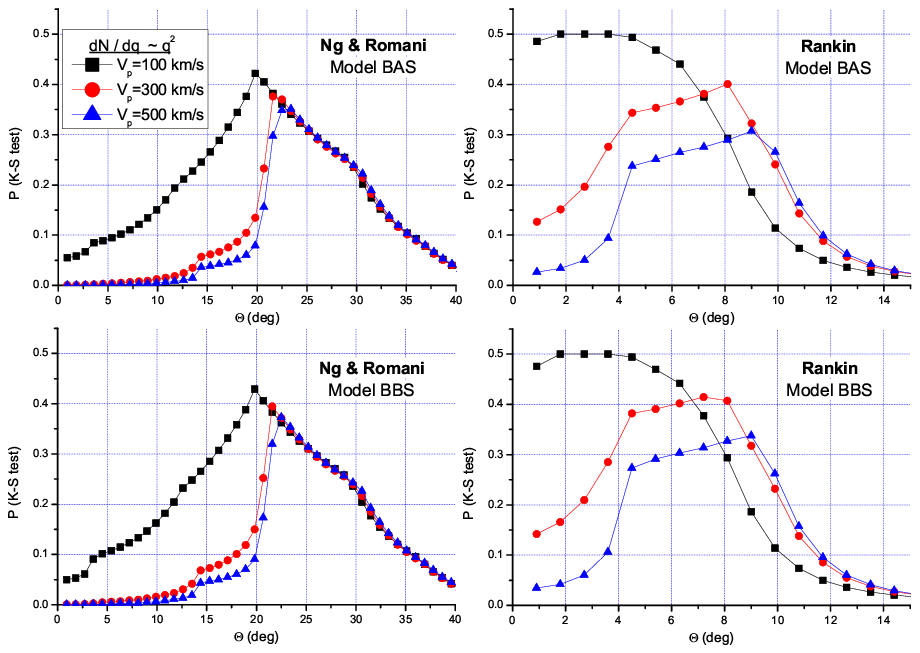}
  \caption{The  Kolmogorov-Smirnov test for the observed 
and calculated spin-velocity alignment angles $\delta$ in pulsars 
as a function of the assumed kick-spin alignment angle $\theta$.
Calculations are shown for the initial binary 
mass ratio distribution $f(q)\sim q^0$ and $f(q)\sim q^2$
(the upper and lower panels, respectively)  
and different models (BAS, BBS).
Squares, circles and triangles mark the kick velocity amplitudes 
100, 300 and 500 km~$s^{-1}$, respectively.  
Left panels: comparison with pulsar data from Ng \& Romani (2007). 
Right panels: comparison with pulsar 
data from Rankin (2007). It is seen that the form of the initial binary mass ratio
distribution and type of kick velocity  models (BAS,BBS) have no influence on the results.
} 
\label{ks_test} 
\end{figure}

The spin-velocity angle distributions are calculated
for different kick-spin angles $\theta$, 
different kick velocities and 
assumptions about the initial binary mass ratio distribution. 
The comparison is made   
with both data sets by \citep{nr2007,r2007}. 
The Kolmogorov-Smirnov (K-S)
test was applied to test the nul hypothesis that the observed 
(Fig.~\ref{obs_bin}) and calculated alignment angles are realizations of
one distribution. 
The results are shown in Figs.~\ref{ks3d_bin_s} -- \ref{ks_test}.
The calculated and observed spin-velocity
angles $\delta$ were binned in 23 two-degree intervals to
increase the number of bins above 20, 
as recommended for the K-S test. 
Since the observational data is folded back to $\delta < 45^\circ$ 
the same folding was applied to the calculated alignment angles.

Fig. \ref{ks3d_bin_s} shows a 3D graph for the
K-S test of the model distribution as a function of the fraction of binary
progenitors (we mix models SA and BB) 
and the kick-spin angle $\theta$. Both panels of Fig.
\ref{ks3d_bin_s} are shown for $v_\mathrm{p}=300$~km~s$^{-1}$ and $dN/dq
\sim q^2$. The left and right panels present comparison  with data from
\cite{nr2007} and from \cite{r2007}, respectively. 
It is visible that better fits are obtained for a high fraction
of binary systems. Especially, in the case of comparison with the data from
\cite{r2007}, small fraction of binary progenitors is disfavoured.

We studied the influence of the fraction of binary progenitors for different
kick velocities on the quality of the K-S test. For both samples of observed
pulsars (from Ng \& Romani and from Rankin) for reasonably high velocities
($v_\mathrm{p}>100$~km~s$^{-1}$) the fraction of binary progenitors should 
be higher than $\sim$50\% to produce high quality of the test. 
For more realistic velocities ($v_\mathrm{p}\sim
200-300$~km~s$^{-1}$) comparison of our calculations with observational data
favours a high fraction of binary progenitors $\ga$70\%, in correspondence
with observations of binary frequencies among massive stars 
\citep{mason98, gies08}.

Fig. \ref{ks_bin_s} shows the K-S test calculated  for
model SA (upper panels) and BB (lower panels), as a function of 
angle $\theta$, for pulsars from \cite{nr2007} (to the left) and
\cite{r2007} (to the right). Fig. \ref{ks_test} shows the K-S test for 
models BAS and 
BBS with different assumptions about the 
initial mass ratio distribution $f(q)$. 

It is seen from Fig. \ref{ks_bin_s} -- \ref{ks_test} that 
the K-S test in all cases does not show good correspondence 
between the observed and calculated pulsar spin-velocity
distributions, which may be related to a poor statistics of pulsars with known $\delta$. 
Nevertheless, the K-S maxima are 
clearly distinguished in all plots, so
the KS-test can be used as a guide to assess the 
relative goodness of the fits.

For all  kick models and initial mass ratio distributions,
the data taken 
from \cite{nr2007} are best-fitted by   
kick velocities confined within the angle  
$\theta\sim 20^\circ$. In contrast, the comparison with 
a more numerous data
by \cite{r2007} favours a narrower natal kick-spin alignment with 
$\theta\sim 5-10^\circ$. This explains our choice of 
the fiducial value $\theta=8^\circ$ we used in Figs.~\ref{new_d}, \ref{psrtypes}. 
Note also that data by \cite{r2007}
can not be fitted by single progenitors only (model SA, right upper
panel in Fig. \ref{ks_bin_s}).

Fig.~\ref{ks_test} clearly shows that 
the results only slightly 
depend on the chosen kick model and the kick
velocity amplitude, and are virtually independent of the assumed initial 
binary mass ratio distribution $f(q)$.

%\begin{figure}
%\includegraphics[width=320pt,angle=0]{bs_frac_test.eps}
%\caption{The value of the ratio of 
%binary and (binary + single) progenitors of the
%population of isolated NSs (in terms of N(model BB)/(N(model BB)+N(model SA))
%whiÓh correspond to maxima magnitude of the
%Kolmogorov-Smirnov test for the observed
%and calculated spin-velocity alignment angles $\delta$ in pulsars.
%Calculations are shown for the model BBS, initial binary
%mass ratio distribution $f(q)\sim q^2$ 
%}
%\label{bin_s_vel_test}
%\end{figure}

It may seem  that the actual shape of 
the kick velocity distribution can be an important parameter. In our 
study we restrict ourselves to considering only the simplest case of a 
single-peak Maxwellian distribution for most of NSs
is in agreement with observations.
As our kick model B assumes the formation of a certain fraction of low-kick NSs and,
hence, low-velocity pulsars, the obtained insignificant
dependence of the calculated spin-velocity alignment on kick models
(at least with the current pulsar data at hands) suggests the true shape of the kick being not so important. 
However, it does matter for the estimation of the relative
fraction of single and binary pulsar progenitors at given $\delta$ and $v$
(Fig. \ref{new_d}). For example, if the true kicks can take values close to zero even 
for single progenitors, it would be hard to distinguish 
binary and single progenitors from such an analysis.

\newpage

\section{Conclusions}

In the present paper we considered the role
of binary progenitors of neutron stars in the apparent
distribution of space velocities and spin-velocity alignment  
of young pulsars. We used a Monte-Carlo synthesis of
pulsar population from single and binary stars under different assumptions 
about the NS natal kick model (form, amplitude and possible
reduction of the kick in binary porgenitors with masses from the range 
8-11 $M_\odot$) and initial binary mass ratio distributions. 
We compared the calculated spin-velocity alignment 
distributions with observational data obtained from radio polarization measurements. 
We arrived at the following conclusions:

1).  The observed space velocity of pulsars is mostly shaped
by the assumed natal kick velocity form and amplitude (Fig. \ref{profile}). 
The effect of binary progenitors is negligible.

2).  The possible kick-spin alignment during the formation of a NS 
strongly affects the spin-velocity correlation in pulsars. 
For the kick model considered,
single progenitors are more
probable for pulsars with space velocities $>50$~km~s$^{-1}$ and tight 
spin-velocity alignment ($<10^\circ$) (Fig. \ref{new_d}).
Binary progenitors  are favoured for pulsars with  
$\delta \ga 15^\circ$ and $v\ga
100$~km~s$^{-1}$, and for low velocity pulsars ($<30$~km~s$^{-1}$) 
for all values of $\delta$.  

3). The comparison with current measurements of pulsar spin-velocity 
angles \citep{nr2007,r2007} does not 
allow us to distinguish between different kick models -- the agreement between
the calculated spin-velocity alignment and the observed distributions
is not very good (Fig. \ref{ks_bin_s}, \ref{ks_test}). 
Nevertheless, the assumption of
natal kick-spin alignment during NS formation
appears to be more important than 
the kick velocity distribution model.

Single progenitors only (model SA, Fig. \ref{ks_bin_s}) do not fit 
data by \cite{r2007} 
and provide worse fit for data 
by \cite{nr2007} than binary progenitors.
Binary progenitors only (model BB, Fig. \ref{ks_bin_s})
can equally well fit both data 
by \cite{nr2007} and \cite{r2007} for the kick-spin 
velocity angle $\theta\sim 5-20^\circ$. 
Models with equal numbers of single and binary progenitors of 
isolated young NSs (models BAS, BBS) equally well fit
observations irrespective of 
the assumed form of the initial binary mass ratio
distribution $f(q)$ (Fig. \ref{ks_test}).
Models without binary progenitors fit observations significantly worse
for standard mean kick velocity amplitude $v_\mathrm{p}\sim
300$~km~s$^{-1}$.

\section*{Acknowledgments} 
The work 
was partially supported by the RFBR grants 06-02-16025, 07-02-00961
and RNP-2.1.1-5940.
S.P. is the INTAS Fellow. S.P. thanks the observatory of Cagliari for 
support and hospitality, special thanks to 
Marta Burgay, Salvatore Pilloni and Andrea Possenti. 
 We also thank the referee for very constructive notes.


\newpage

\begin{thebibliography}{99}
\bibitem[\protect\citeauthoryear{Akiyama et al.}{2003}]{a2003}
Akiyama S. et al., 2003, ApJ, 564, 954

\bibitem[\protect\citeauthoryear{Ardeljan, Bisnovatyi-Kogan \& Moiseenko}{2005}]{abkm05}
Ardeljan N.V., Bisnovatyi-Kogan G.S., Moiseenko S.G., 2005,
MNRAS, 359, 333
\bibitem[\protect\citeauthoryear{Arzoumanian, Chernoff \& Cordes}{2002}]{acc2002}
Arzoumanian Z., Chernoff D.F., Cordes J.M., 2002, ApJ, 568 289
\bibitem[\protect\citeauthoryear{Bassa et al.}{2008}]{40yrs}
Bassa C.G., Wang Z., Cummin A., Kaspi V.M. (Eds.), 2008,
``40 Years of Pulsars: Millisecond Pulsars, Magnetars and More'',
AIP Conf. Proc., vol. 983, American Institute of Physics, Melville, New York   
\bibitem[\protect\citeauthoryear{Berezinsky et al.}{1988}]{b1988}
Berezinskii V.S., Castagnoli C., Dokuchaev V.I., Galeotti P., 1988, Nouvo Com. C, 11, 287 

\bibitem[\protect\citeauthoryear{Bisnovatyi-Kogan}{1993}]{BK1993}
Bisnovatyi-Kogan G.S., 1993,
Astr. Astrophys. Trans., 3, 287

\bibitem[\protect\citeauthoryear{Bombaci \& Popov}{2004}]{bp2004}
Bombaci I., Popov S.B., 2004, A\&A, 424, 627
\bibitem[\protect\citeauthoryear{Chugai}{1984}]{c1984}
Chugai N.N., 1984, Sov. Astron. Lett., 10, 87
\bibitem[\protect\citeauthoryear{Colpi \& Wasserman}{2002}]{cw2002}
Colpi M., Wasserman I., 2002, ApJ, 581, 1271
\bibitem[\protect\citeauthoryear{Gies}{2008}]{gies08}
Gies D.R., 2008, 
	``Massive Star Formation: Observations Confront Theory'', 
ASP Conf. Ser., vol. 387, Eds. Henrik Beuther,
Hendrik Linz, and Thomas Henning, San Francisco: Astron. Soc.
Pac.,  p.93
\bibitem[\protect\citeauthoryear{Grishchuk et al.} {2001}]{Grishchuk}
Grishchuk L.P., Lipunov V.M., Postnov K.A., Prokhorov M.E., Sathyaprakash S., 2001, 
Physics-Uspekhi, 44, 1
[arXiv:astro-ph/0008481]
\bibitem[\protect\citeauthoryear{Faucher-Gigu\'ere \& Kaspi}{2006}]{fk2006}
Faucher-Gigu\'ere C-A., Kaspi V.M., 2006, ApJ, 643,332
\bibitem[\protect\citeauthoryear{Harrison \& Tademaru}{1975}]{ht1975}
Harrison E.R., Tademaru E., 1975, ApJ, 201, 447
\bibitem[\protect\citeauthoryear{Heger et al.}{2005}]{heger&2005}
Heger A., Woosley S., Spruit H.C., 2005, ApJ, 626, 350
\bibitem[\protect\citeauthoryear{Hills} {1983}]{Hills83}
Hills J. G., 1983, ApJ, 267, 322
\bibitem[\protect\citeauthoryear{Hobbs et al.}{2005}]{hobbs2005}
Hobbs G., Lorimer D.R., Lyne A.G., Kramer M., 2005, MNRAS, 360, 974
\bibitem[\protect\citeauthoryear{Iben \& Tutukov}{1996}]{it1996}
Iben I., Tutukov A.V., 1996, ApJ, 456, 738
\bibitem[\protect\citeauthoryear{Imshennik}{1992}]{i1992}
Imshennik V.S., 1992, PAZh, 18, 489
\bibitem[\protect\citeauthoryear{Imshennik \& Nadezhin}{1992}]{in1992}
Imshennik V.S., Nadezhin D.K., 1992, PAZh, 18, 79
\bibitem[\protect\citeauthoryear{Johnston et al.}{2005}]{j2005}
Johnston J., Hobbs G., Vigeland S., Kramer M., Lyne A.G., 2005, MNRAS, 364, 1397
\bibitem[\protect\citeauthoryear{Johnston et al.}{2007}]{j2007}
Johnston J., Kramer M., Karasterigiou A., Hobbs G., Ord S., Wallman J., 2007, 
MNRAS, 381, 1625
\bibitem[\protect\citeauthoryear{Kalogera} {2000}]{Kalogera}
Kalogera V., 2000, ApJ, 541, 319
\bibitem[\protect\citeauthoryear{Khokhlov et al.}{1999}]{k1999}
Khokhlov A. et al., 1999, ApJ, 524, L107 
\bibitem[\protect\citeauthoryear{Lai, Chernoff \& Cordes}{2001}]{lcc2001}
Lai D., Chernoff D.F., Cordes J.M., 2001, ApJ, 549, 1111
\bibitem[\protect\citeauthoryear{Lequeux}{1979}]{l1979}
Lequeux J., 1979, A\&A, 80, 35
\bibitem[\protect\citeauthoryear{Lipunov, Postnov \& Prokhorov}{1996}]{lpp1996}
Lipunov V.M., Postnov K.A., Prokhorov M.E., 1996,  Astroph. Space Sci. Rev.,
9, 1 
\bibitem[\protect\citeauthoryear{Lipunov et al.}{2007}]{lppb2007}
Lipunov V.M., Postnov K.A., Prokhorov M.E., Bogomazov A.I., 2007, arXiv:
0704.1387
\bibitem[\protect\citeauthoryear{Mason et al.}{1998}]{mason98}
Mason B.D., Gies D.R., Hartkopf W.I. et al., 1998, AJ, 115, 821
\bibitem[\protect\citeauthoryear{Ng \& Romani}{2004}]{nr2004}
Ng C.Y., Romani R.W., 2004, ApJ, 601, 479
\bibitem[\protect\citeauthoryear{Ng \& Romani}{2007}]{nr2007}
Ng C.Y., Romani R.W., 2007, ApJ, 660, 1357 
\bibitem[\protect\citeauthoryear{Pinsonneault \& Stanek}{2006}]{ps2006}
Pinsonneault M.H., Stanek K.Z., 2006, ApJ, 639, L67
\bibitem[\protect\citeauthoryear{Podsiadlowski et al.}{2004}]{petal2004}
Podsiadlowski P., Langer N., Poelarends A.J.T., et al., 2004, ApJ, 612, 1044
\bibitem[\protect\citeauthoryear{Poelarends et al.}{2008}]{poletal2008}
Poelarends A.J.T., Herwig F., Langer N., Heger A., 2008, ApJ, 675, 614
\bibitem[\protect\citeauthoryear{Popov \& Prokhorov}{2007}]{pp2007}
Popov S.B., Prokhorov M.E., 2007,  Physics Uspekhi,  50, 1123
\bibitem[\protect\citeauthoryear{Popov} {2008}]{popov2008}
Popov S.B., 2008, Physics of particles and nuclei, 39, 1136
[arXiv:astro-ph/0610593]
\bibitem[\protect\citeauthoryear{Postnov \& Prokhorov} {1998}]{Postnov&Prokhorov}
Postnov K.A., Prokhorov M.E., 1998, Astron. Lett., 24, 586
\bibitem[\protect\citeauthoryear{Postnov \& Yungelson} {2006}]{pyu07}
Postnov K.A., Yungelson L.R., 2006, LRR, 9, 6
\bibitem[\protect\citeauthoryear{Postnov \& Kuranov} {2008}]{postnov&kuranov2008}
Postnov K.A., Kuranov A.G., 2008, MNRAS, 384, 1393
\bibitem[\protect\citeauthoryear{Rankin}{2007}]{r2007}
Rankin J., 2007, ApJ, 664, 443
\bibitem[\protect\citeauthoryear{Scheck et al.}{2004}]{setal2004}
Scheck L., Plewa T., Janka H-Th., Kifonidis K., M\"uller E., 2004,
Phys. Rev. Lett., 92, 011103
\bibitem[\protect\citeauthoryear{Scheck et al.}{2006}]{setal2006}
Scheck L., Kifonidis K., Janka H-Th.,  M\"uller E., 2006, A\&A, 457, 963
\bibitem[\protect\citeauthoryear{Shklovskii}{1970}]{s1970}
Shklovskii I.S., 1970, Sov. Astron., 13, 562
\bibitem[\protect\citeauthoryear{Spruit \& Phinney} {1998}]{Spruit}
Spruit H.C., Phinney E.S., 1998, Nature, 393, 139
\bibitem[\protect\citeauthoryear{Stone}{1991}]{s1991}
Stone R.C., 1991, AJ, 102, 333
\bibitem[\protect\citeauthoryear{Tutukov, Chugai \& Yungelson}{1984}]{tcy1984}
Tutukov A.V., Chugai N.N., Yungelson L.R., 1984, SvA Letters, 10, 244 
\bibitem[\protect\citeauthoryear{van den Heuvel}{2004}]{vdH04}
van den Heuvel E. P. J., 2004, in Proc. 5th Integral Science Workshop, ed.
V. Sch\"onfelder, G. Lichti, \& C. Winkler (ESA SP-552; Noordwijk), p. 185
\bibitem[\protect\citeauthoryear{van den Heuvel} {2007}]{vdH07}
van den Heuvel E.P.J., 2007, in Proc. ``The multicolored landscape of
compact objects and their explosive origins'', AIP Conf. Proc.,
Vol. 924, p. 598 [arXiv:0704.1215]
\bibitem[\protect\citeauthoryear{Wang, Lai \& Han}{2006}]{wlh2006}
Wang C., Lai D., Han J.L., 2006, ApJ, 639, 1007 
\bibitem[\protect\citeauthoryear{Wang, Lai \& Han} {2007}]{Wang_ea07}
Wang C., Lai D., Han J.L., 2007, ApJ, 656, 399
\bibitem[\protect\citeauthoryear{Yakovlev \& Pethick}{2004}]{yp04}
Yakovlev D.G., Pethick C.J., 2004,  ARAA, 42, 169
\bibitem[\protect\citeauthoryear{Zahn}{1977}]{Zahn_77}
Zahn J.-P., 1977, A\&A, 57, 383


\bibitem[\protect\citeauthoryear{}{}]{}
\bibitem[\protect\citeauthoryear{}{}]{}

\end{thebibliography}
\end{document}